\documentclass[lettersize,journal]{IEEEtran}

\usepackage{amsmath,amsfonts,amssymb}
\usepackage{subcaption}
\usepackage{algorithmic}
\usepackage{algorithm}
\usepackage{array}
\usepackage[caption=false,font=normalsize,labelfont=sf,textfont=sf]{subfig}
\usepackage{textcomp}
\usepackage{stfloats}
\usepackage{url}
\usepackage{verbatim}
\usepackage{graphicx}
\usepackage{cite}
\usepackage{pgfplots}
\usepackage{xcolor}
\usepackage{multirow}
\usepackage{makecell}
\usepackage{siunitx}
\usepackage{tabularx}
\usepackage{tabulary}
\usepackage[table]{xcolor}
\definecolor{p}{RGB}{175,141,195}
\definecolor{g}{RGB}{127,191,123}
\definecolor{tabgrey}{RGB}{224,224,224}

\usepackage[utf8]{inputenc}
\usepackage[T1]{fontenc}
\usepackage{cuted}
\usepackage{lipsum}
\usepackage{mathtools}
\usepackage{empheq}

\usetikzlibrary{external}
\usepackage{pgfplots}
\usepackage{tikz}
\usepackage{booktabs}
\usetikzlibrary{patterns}
\usetikzlibrary{decorations.pathreplacing,calligraphy}
\usetikzlibrary{decorations.markings}
\usetikzlibrary{arrows, arrows.meta}
\usepgfplotslibrary{colorbrewer,fillbetween,groupplots}
\usepackage{xcolor}

\definecolor{grad-1}{HTML}{C7B3CC}
\definecolor{grad-2}{HTML}{A7ABC7}
\definecolor{grad-3}{HTML}{87A3C2}
\definecolor{grad-4}{HTML}{669ABC}
\definecolor{grad-5}{HTML}{4692B7}
\definecolor{grad-6}{HTML}{268AB2}

\definecolor{green1}{RGB}{27,158,119}
\definecolor{orange1}{RGB}{217,95,2}
\definecolor{purple1}{RGB}{117,112,179}
\definecolor{pink1}{RGB}{231,41,138}
\definecolor{gray1}{RGB}{102,102,102}
\definecolor{blue1}{RGB}{31,120,180}
\definecolor{pink1}{RGB}{231,41,138}
\definecolor{lightgreen1}{RGB}{102,166,30}

\definecolor{o-band}{RGB}{231,41,138}
\definecolor{e-band}{RGB}{102,166,30}
\definecolor{s-band}{RGB}{117,112,179}
\definecolor{c-band}{RGB}{217,95,2}
\definecolor{l-band}{RGB}{27,158,119}
\usetikzlibrary{shapes.geometric}

\pgfplotsset{compat=1.18}

\usepgfplotslibrary{colorbrewer}
\usepgfplotslibrary{groupplots}
\usepgfplotslibrary[statistics]
\usepgfplotslibrary{fillbetween,groupplots}
\usetikzlibrary{arrows, arrows.meta}
\usetikzlibrary{shapes.geometric}
\newcolumntype{?}{!{\vrule width 2\arrayrulewidth}}

\begin{document}

\title{Energy-Efficient Hollow-Core Fibre Transmission}

\author{Ronit~Sohanpal, Eric~Sillekens, Mindaugas~Jarmolovi\v{c}ius, Robert~I.~Killey and Polina~Bayvel

\thanks{Manuscript received XX XX, XXXX; revised XX XX, XXXX.}%
\thanks{This work was supported by EPSRC grants EP/R035342/1 Transforming Networks - building an intelligent optical infrastructure (TRANSNET) and EP/W015714/1 Extremely Wideband Optical Fibre Communication Systems (EWOC)). DBIST (supporting ES) is gratefully acknowledged. \emph{(Corresponding author: Ronit Sohanpal.)}}%
\thanks{Ronit~Sohanpal, Eric~Sillekens, Mindaugas~Jarmolovi\v{c}ius, Robert~I.~Killey and Polina~Bayvel are with the Optical Networks Group, Department of Electronic and Electrical Engineering, UCL (University College London), London, UK (E-mail: ronit.sohanpal@ucl.ac.uk).} %
}

\markboth{Journal of XXXX,~Vol.~XX, No.~XX, XX~XXXX}%
{XXXX \MakeLowercase{\textit{et al.}}: Bare Demo of IEEEtran.cls for IEEE Journals}

\maketitle

\begin{abstract}
Hollow-core fibres (HCFs) are a promising means of increasing the throughput of coherent transmission systems. In addition to their advantages in terms of low latency, nonlinearity and attenuation, HCFs can potentially improve the energy efficiency of coherent transmission systems by reducing the number of repeaters and enabling more efficient modulation formats than SMF links. However, the relationship between the link parameters (e.g. launch power, amplifier efficiency and transceiver noise) and the energy efficiency has not been explored.

In this work, we investigate energy-efficient operating regimes in HCF transmission systems. We show that the optimum energy per bit in SMF systems is ultimately throughput-limited - maximising throughput will minimise energy per bit. In contrast, the transceiver-limited throughput of HCF leads to two separate launch power optima - minimum-energy-per-bit and maximum-throughput. We derive a closed-form equation for the minimum-energy-per-bit launch power for HCF links in terms of the link parameters, including the amplifier efficiency, transceiver power consumption and link gain. We use our model to explore the impact of span length and fibre attenuation in both operating regimes, showing how energy per bit considerations significantly impact the optimum span length. Optimising for energy efficiency can lead to 50\% reduction in link energy per bit for only a 3\% throughput penalty at 3000~km, whilst also reducing the required amplifier launch power from \textgreater~33~dBm to \textless~23~dBm. This work highlights the importance of including physical layer energy considerations in HCF link design.

\end{abstract}


\section{Introduction}

Hollow-core fibres (HCFs) have seen significant research interest in recent years as a means to enhance the performance of coherent optical fibre communication links. HCFs possess an air-filled core that enables numerous advantages over standard silica solid-core single-mode fibres (SMFs), including a 30\% reduction in latency, three to four orders of magnitude reduction in nonlinearity, several orders of magnitude reduction in the Rayleigh backscattering coefficient and a reduction in attenuation below 0.1~dB/km \cite{petrovichBroadbandOpticalFibre2025,Penglowestloss,gao2025}. These advancements have made possible a variety of novel transmission capabilities and record experiments that were unachievable in SMF. Notable examples include \textgreater~300~km unrepeatered transmission \cite{Ali2025}, real-time \textgreater 10,000~km recirculating-loop transmission \cite{Hong2025longhaul}, \textgreater 222~km HCF-SMF hybrid-span systems \cite{mardoyan222kmlongHybridSpan2026} and \textgreater 400~Tb/s OESCL-band bidirectional transmission \cite{yang2026423}.

An interesting aspect of HCF deployment not yet deeply investigated is its impact on optical link energy efficiency. Longer span lengths can reduce the number of required repeaters for fixed-length links, leading to a net reduction in the total amplifier power consumption \cite{Sticca26,sticcaHighPowerOpticalAmplification2024}. In addition, the higher achievable signal-to-noise ratio (SNR) of an HCF link versus an equivalent SMF link means it can achieve higher energy efficiencies than SMF systems despite transmitting at higher launch powers. However, the energy efficiency of HCF systems has only been studied in a network context, with the impact of various link parameters including launch power, amplifier efficiency and transceiver noise on the net energy efficiency still not understood.

In this work, we investigate the energy efficiency of single-mode  fibre and hollow-core fibre links. This work is an extension of \cite{sohanpal2026optimum}. We show that in C-band SMF systems, the most energy-efficient launch power (i.e. that which minimises the energy per bit) is the one that maximises throughput. In contrast, we show that HCF systems are limited by transceiver noise rather than nonlinearity, leading to two separate launch power operation regimes - maximum-throughput and minimum-energy-per-bit. We derive a closed-form solution to the minimum-energy-per-bit in HCF systems, verify its accuracy and investigate the impact of HCF attenuation and span length on the optimum link design.
\section{Modelling Energy-per-bit}

The total electrical power consumption of a coherent WDM link can be calculated as the sum of the total electrical power consumed by the amplifiers and the transceivers. The total electrical power consumption of a doped-fibre amplifier (DFA) for a transparent $N_{\text{span}}$-span link is \cite{sohanpal2026raman}
\begin{equation} \label{eq:pamp}
    P_{\text{amp}} =   \frac{N_{\text{span}}}{\eta_{\text{PCE}}} \Big(1-\frac{1}{G}\Big)P_{\text{out}} + N_{\text{span}}P_{\text{mm}}
\end{equation}

where $P_{\text{out}}$ is the total output power of the DFA, $G$ is the amplifier gain, $P_{\text{mm}}$ is the per-amplifier monitoring and management overhead power consumption and $\eta_{\text{PCE}}$ is the amplifier wallplug power conversion efficiency (PCE), which reflects the net energy efficiency of all the conversion processes within the amplifier (e.g. driver electrical-to-electrical conversion, pump electrical-to-optical conversion, doped-fibre optical-to-optical conversion etc). For most C-band links, the amplifier gain $G$ is large (\textgreater15~dB), therefore $P_{\text{out}} (1-1/G)\approx P_{\text{out}}$.

The total electrical power consumption of the transceivers $P_{\text{TRX}}$ is the sum of the electrical power of each channel, treated here as launch-power-independent. This assumption neglects the wavelength-dependence of the transceiver subsystems (e.g. the laser, the chromatic dispersion compensation block) as well as the signal-quality-dependence of the DSP (including the FEC), but determination of these parameters is complex and depends on the transceiver material platform, DSP architecture and laser technology, and have not been quantified in literature.

The output power $P_{\text{out}}$ can be rewritten as $\sum_{i=1}^{N_{\text{ch}}} P_i$ where $N_{\text{ch}}$ is the total number of channels and $P_i$ is the launch power per channel. The total electrical power consumption is therefore
\begin{equation}\label{eq:electrical}
    P_{\text{elec}} = \frac{N_{\text{span}}}{\eta_{\text{PCE}}}\sum_{i=1}^{N_{\text{ch}}} \bigg((1-1/G_i) P_i\bigg) + N_{\text{ch}}P_{\text{TRX}} + N_{\text{span}}P_{\text{mm}}
\end{equation}

The total throughput can be determined from the Shannon capacity formula, assuming additive white Gaussian noise (AWGN) and ideal matched filtering
\begin{equation} \label{eq:throughput}
    T = \sum_{i=1}^{N_{\text{ch}}} 2 R \log_2(1 + \text{SNR}_i) \\
\end{equation}

where $R$ is the channel baud rate and
\begin{equation} \label{eq:snrtot}
    \text{SNR}_i = \frac{P_i}{P_{\text{ASE}} + \eta_{\text{NLI}}P_i^3 + \kappa P_i}
\end{equation}

where $P_{\text{ASE}}$ is the accumulated ASE noise power across all spans and $\eta_{\text{NLI}}$ is the nonlinear interference parameter due to fibre nonlinearity. Here, $\kappa P_i$ represents the power-independent noise terms, namely the combined transceiver noise and inter-modal interference noise (IMI). In SMF, IMI is negligible, thus $\kappa$ reflects only the unitless transceiver noise coefficient $\kappa = \kappa_{\text{TRX}}$. In HCF, IMI is non-negligible for long transmission distances, thus $\kappa$ becomes
\begin{equation}
    \kappa = \kappa_{\text{TRX}} + \kappa_{\text{IMI}} L_{\text{tot}}
\end{equation}

where $\kappa_{\text{IMI}}$ is the IMI coefficient in units of $1/\text{km}$ and $L_{\text{tot}}$ is the total transmission distance. The maximum throughput $T_{\text{max}}$ occurs at maximum SNR. The optimum-throughput launch power $P_{\text{opt}}^T$ is
\begin{align} \label{eq:maxthroughput}
P_{\text{opt}}^T = \sqrt[3]{\frac{P_{\text{ASE}}}{2\eta_{\text{NLI}}}} ,\ \text{SNR}_{\text{max}} = \frac{P_{\text{opt}}^T}{\frac{3}{2}P_{\text{ASE}} + \kappa P_{\text{opt}}^T}
\end{align}

Combining (\ref{eq:electrical}) and (\ref{eq:throughput}), the total link energy-per-bit $E_b$ is
\begin{align} 
    E_b &= \frac{P_{\text{elec}}}{T}\\
    &= \frac{ \frac{N_{\text{span}}}{\eta_{\text{PCE}}} \sum_{i=1}^{N_{\text{ch}}} \big((1-1/G_i)P_i \big) + N_{\text{ch}}P_{\text{TRX}} + N_{\text{span}}P_{\text{mm}}}{ \sum_{i=1}^{N_{\text{ch}}}2 R \log_2(1 + \text{SNR}_i)} \label{eq:generalform}
\end{align}

Eq.~(\ref{eq:generalform}) can be solved using either the integral or closed-form Gaussian Noise (GN) models to determine $\eta_{\text{NLI}}$ and thus determine $E_b$. The most energy-efficient launch power is the one that minimises $E_b$, and is represented here as $P_{\text{opt}}^{E_b}$. Due to the logarithm in the denominator there is no closed-form solution for the minimum, but suitable approximated solutions can be found depending on the fibre and operating regime.

\section{Methodology}
\label{sec:methodology}

To investigate the energy-per-bit numerically, the closed-form GN model was used to estimate the per-channel SNR \cite{poggioliniOpportunitiesChallengesLongDistance2022}. The noise is assumed to be additive, white and Gaussian; the transceiver noise, ASE, IMI and NLI where applicable are assumed to be independent; inter-channel stimulated Raman scattering (ISRS) is assumed to be negligible in this work. Unlike in our previous work \cite{sohanpal2026optimum}, we focus here on only the C-band (1530 - 1565~nm), justifying the aforementioned neglection of ISRS, but this work can be extended to other bands by appropriately modelling ISRS and other wavelength-dependent parameters \cite{buglia2024}.

The fully-loaded C-band was modelled with 140~GBd dual-polarisation Gaussian-modulation channels with a 150~GHz spacing and transceiver SNR of 20~dB ($\kappa_{\text{TRX}}=$0.01), totalling 29 channels. Two fibres are considered here: HCF and SMF. The HCF span length was 200~km with a dispersion $D = 3.5$~(ps/(nm$\cdot$km), attenuation $\alpha = 0.05$~dB/km, IMI coefficient of -52~dB/km \cite{ospinaLeveragingDigitalSubcarrier2026} and a nonlinear coefficient $\gamma$ of $5{\times}10^{-4}$ W$^{-1}$km$^{-1}$. No gas line absorption was included in the HCF modelled here as it has only minimal effect in the C-band, but can be included as a wavelength-dependent attenuation, and is mitigable through optimised fabrication \cite{xiongTextCO_2EliminationHollowCore2025}. The SMF had a dispersion coefficient $D = 17$~(ps/(nm$\cdot$km), nonlinear coefficient $\gamma = 1.2$~(W$\cdot$km)$^{-1}$, and attenuation $\alpha = 0.2$~dB/km. IMI was assumed to be negligible. The noise figure of the C-band EDFA was 5~dB, and a constant launch power was used for all channels.

A constant value for the PCE was used here, i.e. it is assumed the EDFA was designed to operate at the desired output power, and that the PCE saturates at a steady state at very high launch powers. This assumption means the modelling will overestimate the power efficiency at very low and very high output powers \cite{sohanpal2026raman,Sohanpal26uwb}. Since most in-line EDFAs are designed to operate around 20~dBm saturation, the former will not affect the conclusions of this study. Typically the pump driver efficiency reduces when supplying large currents due to excess thermal dissipation, but the efficiency of EDFAs designed for such high output powers has not been reported in the literature. A PCE value of 5\% was used here for a C-band EDFA \cite{sohanpal2026raman,liangRepeaterPowerConversion2021}.

For each transceiver, a constant 24~W per transceiver was used, corresponding to the approximate power consumption of an 800G pluggable transceiver module with an efficiency of 3~W per 100~Gb/s. A $P_{\text{mm}}$ value of 2~W was used for the monitoring and management power consumption as a reference value, but typically this is small compared to the total transceiver power consumption.

\section{Results}

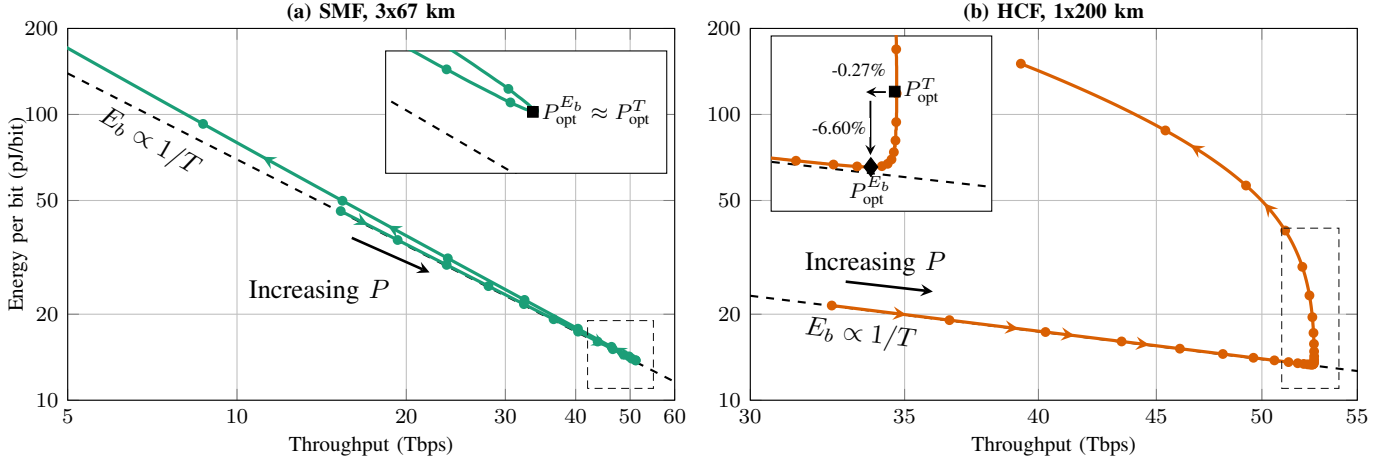
\begin{figure*}[ht!]
\tikzsetnextfilename{EbvsT}
\begin{tikzpicture}[font=\footnotesize]
    \begin{groupplot}
    [
    width=0.53\linewidth,
    height=6.5cm,
    grid=major,
    ylabel near ticks,
    xlabel near ticks,
    ylabel shift = -8 pt,
    xlabel shift = -2 pt,
    clip marker paths=true,
    ymode=log,
    xmode=log,
    xlabel=Throughput (Tbps),
    ylabel=Energy per bit (pJ/bit),
    clip mode=individual,
    label style={font=\footnotesize},
    tick label style={font=\footnotesize},
    legend style={fill opacity=1, draw opacity=1, text opacity=1, at={(0.5,0.97)}, anchor=north west, draw=black, nodes={scale=0.75, transform shape}, legend columns=1},
    group style={group size=2 by 1,xlabels at=edge bottom,ylabels at=edge left,horizontal sep=1cm},
    title style={at={(0.5,0.95)},font=\bfseries\footnotesize},
    ]

    \nextgroupplot[
    domain = 5:60,
    xtick={5,10,20,30,40,50,60},
    xticklabels={$5$,$10$,$20$,$30$,$40$,$50$,$60$},
    ytick = {10,20,50,100,200},
    yticklabels={$10$,$20$,$50$,$100$,$200$},
    ymin=10,
    ymax=200,
    xmin=5,
    xmax=60,
    title={\textbf{(a)} SMF, 3x67~km},
    ]

    \node[rotate=-30] at (axis cs: 7,80) {\normalsize $E_b \propto 1/T $};
    
    \addplot[black,
    no marks, 
    dashed,
    line width = 0.8pt,
    ]plot {(29*24)/\x};    

    \addplot[l-band,
    no marks,  
    line width=1.2pt,
    postaction={decorate, decoration={
            markings,
            mark=at position 0.03 with {\arrow{stealth}},
            mark=at position 0.29 with {\arrow{stealth}},
            mark=at position 0.35 with {\arrow{stealth}},
            mark=at position 0.6 with {\arrow{stealth}},
            mark=at position 0.74 with {\arrow{stealth}},
        }}
    ] table[x=cap,y=pjbit] {data/SMF_EbvsT_3x67.txt};
    
    \addplot[l-band,
    mark=*, 
    only marks,
    mark size = 1.5, 
    line width=0.7pt,
    each nth point=20,
    ] table[x=cap,y=pjbit] {data/SMF_EbvsT_3x67.txt};

    \draw [-stealth,line width = 1] (axis cs: 16,37)--(axis cs:22,28);
    \node[] at (axis cs: 14,24) {\normalsize Increasing $P$};
    
    \draw[black!90,densely dashed] (42,11) rectangle (55,19);

    \nextgroupplot[
     xtick={30,35,40,45,50,55},
    xticklabels={$30$,$35$,$40$,$45$,$50$,$55$},
    ytick = {10,20,50,100,200},
    yticklabels={$10$,$20$,$50$,$100$,$200$},
    ymin=10,
    ymax=200,
    xmin=30,
    xmax=55,
    domain = 30:60,
    title={\textbf{(b)} HCF, 1x200~km},
    ]

    \addplot[black,
    no marks, 
    dashed,
    line width = 0.8pt,
    ]plot {(29*24)/\x};

    \addplot[c-band,
    no marks,  
    line width=1.2pt,
     postaction={decorate, decoration={
            markings,
            mark=at position 0.08 with {\arrow{stealth}},
            mark=at position 0.2 with {\arrow{stealth}},
            mark=at position 0.26 with {\arrow{stealth}},
            mark=at position 0.34 with {\arrow{stealth}},
            mark=at position 0.7 with {\arrow{stealth}},
            mark=at position 0.8 with {\arrow{stealth}},
        }}
    ] table[x=cap,y=pjbit] {data/HCF_EbvsT_1x200.txt};
    
    \addplot[c-band,
    mark=*, 
    only marks,
    mark size = 1.5, 
    line width=0.7pt,
    each nth point=20,
    ] table[x=cap,y=pjbit] {data/HCF_EbvsT_1x200.txt};

    \node[rotate=-7] at (axis cs: 33.5,17) {\normalsize $E_b \propto 1/T $};

    \draw [-stealth,line width = 1] (axis cs: 33,26)--(axis cs:36,24);
    \node[] at (axis cs: 34,30) {\normalsize Increasing $P$};

    \draw[black!90,densely dashed] (51,40) rectangle (54,11);

    \end{groupplot}

    \begin{scope}[shift={(4.2,3)}]
        \begin{axis}
    [
    width=5.3cm,
    height=3.2cm,
    grid=major,
    ylabel near ticks,
    xlabel near ticks,
    ylabel shift = -8 pt,
    xlabel shift = -2 pt,
    clip marker paths=true,
    ymode=log,
    xmode=log,
    clip mode=individual,
    label style={font=\footnotesize},
    tick label style={font=\footnotesize},
    legend style={fill opacity=1, draw opacity=1, text opacity=1, at={(0.1,0.5)}, anchor=north west, draw=black, nodes={scale=1, transform shape}, legend columns=1},
    group style={group size=2 by 1,xlabels at=edge bottom,ylabels at=edge left,horizontal sep=1.2cm},
    domain = 40:53,
    xtick = {\empty},
    ytick = {\empty},
    ymin=13.6,
    ymax=13.9,
    xmin=50.5,
    xmax=52,
    axis background/.style={fill=white}
    ]

    \addplot[black,
    no marks, 
    dashed,
    line width = 0.8pt,
    ]plot {(29*24)/\x};    

    \addplot[l-band,
    no marks,  
    line width=1.2pt,
    postaction={decorate, decoration={
            markings,
            mark=at position 0.15 with {\arrow{stealth}},
        }}
    ] table[x=cap,y=pjbit] {data/SMF_EbvsT_3x67.txt};
    
    \addplot[l-band,
    mark=*, 
    only marks,
    mark size = 1.5, 
    line width=0.7pt,
    each nth point=10,
    ] table[x=cap,y=pjbit] {data/SMF_EbvsT_3x67.txt};

    \addplot[black, only marks,mark=square*] coordinates {(51.28,13.75)};
    \node[right] at (axis cs:51.28,13.75) {$P_{\text{opt}}^{E_b} \approx P_{\text{opt}}^{T}$};

        \end{axis}
    \end{scope}

    \begin{scope}[shift={(9.3,2.5)}]
        \begin{axis}
    [
    width=4.5cm,
    height=3.9cm,
    grid=major,
    ylabel near ticks,
    xlabel near ticks,
    ylabel shift = -8 pt,
    xlabel shift = -2 pt,
    clip marker paths=true,
    ymode=log,
    xmode=log,
    clip mode=individual,
    label style={font=\footnotesize},
    tick label style={font=\footnotesize},
    legend style={fill opacity=1, draw opacity=1, text opacity=1, at={(0.1,0.5)}, anchor=north west, draw=black, nodes={scale=1, transform shape}, legend columns=1},
    group style={group size=2 by 1,xlabels at=edge bottom,ylabels at=edge left,horizontal sep=1.2cm},
    domain = 50:55,
    xtick = {\empty},
    ytick = {\empty},
    ymin=12.8,
    ymax=15,
    xmin=52,
    xmax=53.2,
    axis background/.style={fill=white}
    ]
    
    \addplot[black,
    no marks, 
    dashed,
    line width = 0.8pt,
    ]plot {(29*24)/\x};    

    \addplot[c-band,
    no marks,  
    line width=1.2pt,
     postaction={decorate, decoration={
            markings,
            mark=at position 0.03 with {\arrow{stealth}},
        }}
    ] table[x=cap,y=pjbit] {data/HCF_EbvsT_1x200.txt};
    
    \addplot[c-band,
    mark=*, 
    only marks,
    mark size = 1.5, 
    line width=0.7pt,
    each nth point=20,
    ] table[x=cap,y=pjbit] {data/HCF_EbvsT_1x200.txt};

    \addplot[black, only marks,mark=diamond*,mark size = 3.5pt] coordinates {(52.539,13.323)};
    \node[below] at (axis cs:52.539,13.323) {$P_{\text{opt}}^{E_b}$};

    \addplot[black, only marks,mark=square*] coordinates {(52.67,14.264)};
    \node[right] at (axis cs:52.67,14.264) {$P_{\text{opt}}^{T}$};

    \draw [stealth-,line width = 0.7] (rel axis cs: 0.45,0.32)--(rel axis cs:0.45,0.63) node[midway,fill=white,inner sep=1pt,font=\scriptsize,anchor=east] {-6.60\%};
    
    \draw [stealth-,line width = 0.7] (rel axis cs: 0.43,0.68)--(rel axis cs:0.52,0.68);

    \node[font=\scriptsize] at (rel axis cs: 0.4,0.81) {-0.27\%};

        \end{axis}
    \end{scope}

\end{tikzpicture}

\caption{Energy-per-bit versus throughput for 200~km C-band transmission and increasing per-channel launch power (solid lines) for (a) SMF and (b) HCF. Insets indicate the minimum energy-per-bit (diamond) and maximum throughput (square).}
\label{fig:EbvsT}
\end{figure*}

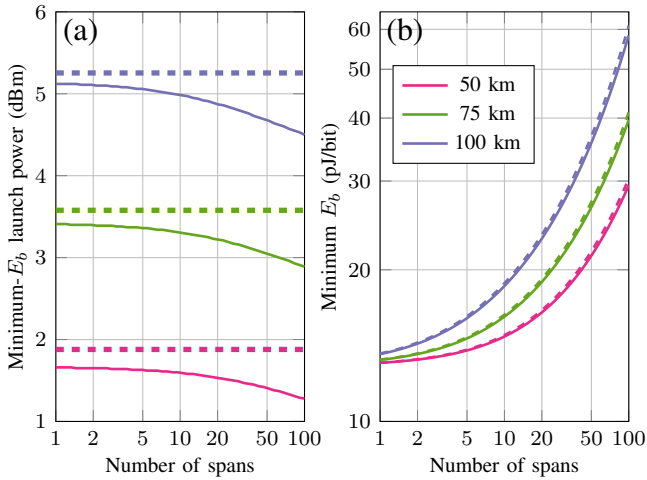
\begin{figure}[ht!]
\tikzsetnextfilename{SMFNspanSweep}
\begin{tikzpicture}[font=\footnotesize]
    \begin{groupplot}
    [
    width=0.55\linewidth,
    height=7cm,
    grid=major,
    ylabel near ticks,
    xlabel near ticks,
    ylabel shift = -4 pt,
    xlabel shift = -2 pt,
    clip marker paths=true,
    xmode=log,
    xlabel=Number of spans,
    clip mode=individual,
    label style={font=\footnotesize},
    tick label style={font=\footnotesize},
    legend style={fill opacity=1, draw opacity=1, text opacity=1, at={(0.05,0.87)}, anchor=north west, draw=black, nodes={scale=1, transform shape}, legend columns=1},
    group style={group size=2 by 1,xlabels at=edge bottom,ylabels at=edge left,xticklabels at=edge bottom,horizontal sep=1cm},
    title style={at={(0.5,0.95)},font=\bfseries\footnotesize},
    xmin=1,
    xmax=100,
    xtick={1,2,5,10,20,50,100},
    xticklabels={$1$,$2$,$5$,$10$,$20$,$50$,$100$},
    ]

    \nextgroupplot[
    ymin=1,
    ymax=6,
    ylabel={Minimum-$E_b$ launch power (dBm)},
    ytick={1,2,3,4,5,6},
    ]

    \addplot[domain=1:100, o-band, line width=2pt, dashed] {1.8789};
    \addplot[domain=1:100, e-band, line width=2pt, dashed] {3.5776};
    \addplot[domain=1:100, s-band, line width=2pt, dashed] {5.2542};

    \addplot[o-band,
    no marks,
    line width=1pt,
    ] table[x=nspans,y=popt50] {data/SMF_Nspansweep.txt};

    \addplot[e-band,
    no marks,
    line width=1pt,
    ] table[x=nspans,y=popt75] {data/SMF_Nspansweep.txt};

    \addplot[s-band,
    no marks,
    line width=1pt,
    ] table[x=nspans,y=popt100] {data/SMF_Nspansweep.txt};

    \node at (rel axis cs: 0.1, 0.95) {\large (a)};

    \nextgroupplot[
    ymode = log,
    ymin=10,
    ymax=65,
    ytick={10,20,30,40,50,60},
    yticklabels={$10$,$20$,$30$,$40$,$50$,$60$},
    ylabel=Minimum $E_b$ (pJ/bit),
    ]

    \addplot[o-band,
    no marks,
    line width=1pt,
    ] table[x=nspans,y=eb50] {data/SMF_Nspansweep.txt};
    \legend{50 km, 75 km, 100 km}

    \addplot[e-band,
    no marks,
    line width=1pt,
    ] table[x=nspans,y=eb75] {data/SMF_Nspansweep.txt};

    \addplot[s-band,
    no marks,
    line width=1pt,
    ] table[x=nspans,y=eb100] {data/SMF_Nspansweep.txt};

    \addplot[o-band,
    no marks, dashed,
    line width=1pt,
    ] table[x=nspans,y=eb50theory] {data/SMF_Nspansweep.txt};

    \addplot[e-band,
    no marks, dashed,
    line width=1pt,
    ] table[x=nspans,y=eb75theory] {data/SMF_Nspansweep.txt};

    \addplot[s-band,
    no marks, dashed,
    line width=1pt,
    ] table[x=nspans,y=eb100theory] {data/SMF_Nspansweep.txt};

    \node at (rel axis cs: 0.1, 0.95) {\large (b)};
    
    \end{groupplot}

    \end{tikzpicture}

\caption{(a) Minimum-$E_b$ launch power and (b) minimum $E_b$ versus number of spans for different SMF span lengths. Solid lines indicate numerical results, dashed lines obtained  assuming $P_{\text{opt}}^{E_b} = P_{\text{opt}}^{T} $ using (\ref{eq:maxthroughput}).}
\label{fig:Nspan_smf}
\end{figure}

\subsection{Optimum $E_b$ in SMF}

In SMF transmission systems, the maximum SNR is limited by the fibre nonlinearity. Consider an 800~km fully-loaded C-band link consisting of 80~km spans, with the same parameters as in Section \ref{sec:methodology} and setting $P_{\text{mm}}=0$. The optimum-throughput launch power is $P_{\text{opt}}^{T} = 3.64$~dBm per channel (18.26~dBm total) using (\ref{eq:maxthroughput}). Plugging this into (\ref{eq:pamp}), for $\eta_{\text{PCE}}=5$\% the amplifier power consumption is approximately 13~W. The total transceiver power consumption for 29 channels is 696~W for the entire C-band. Here, the EDFA only contributes approximately 1.84\% of the total electrical power when operating at maximum throughput, thus, in typical SMF transmission systems, the transceivers' power consumption dominates over that of the amplifiers. 
Neglecting the amplifier power consumption term in (\ref{eq:electrical}), the total electrical power will be determined by the launch-power-independent terms. Consequently, the minimum $E_b$ occurs when the throughput is maximised, and the corresponding optimum launch power for SMF is $P_{\text{opt}}^{E_b} \approx P_{\text{opt}}^{T}$ in (\ref{eq:maxthroughput}). For SMF systems, the most energy-efficient launch power is that which maximises throughput.

Fig.~\ref{fig:EbvsT}a shows the total  energy per bit $E_b$ versus throughput $T$ for C-band transmission in a 3x67~km SMF link as the launch power per channel is swept from -20 to 20 dBm. At low launch powers (e.g. below -5~dBm per channel), $E_b \propto 1/T$ as the amplifier power consumption is negligible, thus $E_b$ decreases proportionally to the increase in throughput as the launch power increases. At optimum launch power (3~dBm per channel), the minimum-$E_b$ power is approximately equal to the maximum-$T$ power. Beyond the optimum, the throughput decreases rapidly and $E_b$ increases (but still nearly proportionally to $1/T$), indicating that even for channel powers approaching 20~dBm, the amplifier electrical power contribution is small.

Fig.~\ref{fig:Nspan_smf} shows the optimum launch power $P_{\text{opt}}^{E_b}$ and minimum $E_b$ as a function of number of spans and for different SMF span lengths. The dashed lines are obtained assuming $P_{\text{opt}}^{E_b} = P_{\text{opt}}^{T}$ using (\ref{eq:maxthroughput}) and (\ref{eq:generalform}). For few spans, this assumption shows good agreement with the simulated model. As the span count increases, the amplifier electrical power consumption increases, causing the minimum-$E_b$ launch power to deviate from (\ref{eq:maxthroughput}). Nevertheless, this assumption is valid for these C-band links with 6.2\% error in $P_{\text{opt}}^{E_b}$ corresponding to only 1.3\% error in minimum-$E_b$ at 10 spans for the worst-case 100~km span length.

\subsection{Optimum $E_b$ in HCF}

Fig.~\ref{fig:EbvsT}b shows the total energy per bit $E_b$ versus throughput $T$ for a 1x200~km HCF link as the launch power per channel is swept from -20 to 50 dBm. At low launch powers (e.g. below -5~dBm per channel), this exhibits the same throughput-limited $E_b$ as in the SMF case. However, there is a clear divergence from the SMF case near the optimum launch power, where $P_{\text{opt}}^{E_b} \neq P_{\text{opt}}^{T}$. As the HCF throughput is strongly limited by transceiver noise rather than nonlinearity (as explored in \cite{Klaus2022}), this leads to the throughput increasing slowly relative to the amplifier power consumption as the launch power increases, leading to greatly differing values of $P_{\text{opt}}^{E_b} = 5.8$~dBm and $P_{\text{opt}}^{T} = 20$~dBm per channel. Thus, (\ref{eq:maxthroughput}) does not reflect the most energy-efficient launch power for HCF links. Moving from $P_{\text{opt}}^{T}$ to $P_{\text{opt}}^{E_b}$, the throughput decreases by only 0.27\% while $E_b$ reduces by 6.6\%, thus significant improvements in energy efficiency can be obtained in HCF links at marginal throughput penalties.

Making appropriate assumptions, it is possible to derive a closed-form equation for the minimum-$E_b$ launch power in HCF. First, it is assumed that the nonlinearity is negligible in the vicinity of $P_{\text{opt}}^{E_b}$. Since the throughput is limited by transceiver noise, the very small nonlinearity of HCF can be ignored. Assuming frequency-independent attenuation and dispersion within the transmission bandwidth and zero nonlinearity (i.e. identical and independent channels), (\ref{eq:generalform}) can be rewritten as

\begin{equation} \label{eq:rearrange}
    E_b = \frac{ N_{\text{ch}}N_{\text{span}} (1-1/G)P + \eta_{\text{PCE}} N_{\text{ch}}P_{\text{TRX}} + \eta_{\text{PCE}}N_{\text{span}}P_{\text{mm}} }{ 2 RN_{\text{ch}} \eta_{\text{PCE}} \log_2(1 + \text{SNR})} 
\end{equation}

\begin{figure*}[th!]
    \centering
    \begin{equation} \label{eq:exactclosed}
    P_{\text{opt}}^{E_b} =  \frac{P_{\text{ASE}}}{\kappa}\Bigg(\frac{2(r-\frac{\kappa+2}{\kappa+1})}{-2 - \ln(1 + 1/\kappa) + \sqrt{(2+\ln(1 + 1/\kappa))^2 + 4(r-\frac{\kappa+2}{\kappa+1})(\kappa+1)\ln(1 + 1/\kappa)}}   - 1\Bigg)
    \end{equation}
    \noindent\rule{\textwidth}{0.5pt}
\end{figure*}

Taking the derivative with respect to $P$ and assuming that $P_{\text{ASE}}$ is small compared to the transceiver noise near the optimum (see Appendix) to find the positive root, we can obtain a closed-form solution  (\ref{eq:exactclosed}), where
\begin{equation} \label{eq:r}
    r = \frac{\eta_{\text{PCE}}\kappa (N_{\text{ch}}P_{\text{TRX}} + N_{\text{span}}P_{\text{mm}})}{N_{\text{ch}}N_{\text{span}}P_{\text{ASE}}(1-1/G)}
\end{equation}

\begin{figure}[t]
\tikzsetnextfilename{HCFNspanSweep}
\begin{tikzpicture}[font=\footnotesize]
    \begin{groupplot}
    [
    width=0.55\linewidth,
    height=7cm,
    grid=major,
    ylabel near ticks,
    xlabel near ticks,
    ylabel shift = -7 pt,
    xlabel shift = -2 pt,
    clip marker paths=true,
    xmode=log,
    xlabel=Number of spans,
    clip mode=individual,
    label style={font=\footnotesize},
    tick label style={font=\footnotesize},
    legend style={fill opacity=1, draw opacity=1, text opacity=1, at={(0.05,0.87)}, anchor=north west, draw=black, nodes={scale=1, transform shape}, legend columns=1},
    group style={group size=2 by 1,xlabels at=edge bottom,ylabels at=edge left,xticklabels at=edge bottom,horizontal sep=1cm},
    title style={at={(0.5,0.95)},font=\bfseries\footnotesize},
    xmin=1,
    xmax=100,
    xtick={1,2,5,10,20,50,100},
    xticklabels={$1$,$2$,$5$,$10$,$20$,$50$,$100$},
    ]

    \nextgroupplot[
    ymin=1,
    ymax=11,
    ylabel={Minimum-$E_b$ launch power (dBm)},
    mark size=2pt,
    ytick={1,3,5,7,9,11},
    ]

    \addplot[o-band,
    no marks,
    line width = 1pt,
    ] table[x=nspans,y=popt100] {data/HCF_Nspansweep.txt};

    \addplot[e-band,
    no marks,
    line width = 1pt,
    ] table[x=nspans,y=popt200] {data/HCF_Nspansweep.txt};

    \addplot[s-band,
    no marks,
    line width = 1pt,
    ] table[x=nspans,y=popt300] {data/HCF_Nspansweep.txt};
    
    \addplot[c-band,
    no marks,
    line width = 1pt,
    ] table[x=nspans,y=popt400] {data/HCF_Nspansweep.txt};

    \addplot[o-band,
    mark=square*,
    mark options={fill=white},
    only marks,
    each nth point=4,
    ] table[x=nspans,y=popt100theoryapprox] {data/HCF_Nspansweep.txt};

    \addplot[e-band,
    mark=square*,
    mark options={fill=white},
    only marks,
    each nth point=4,
    ] table[x=nspans,y=popt200theoryapprox] {data/HCF_Nspansweep.txt};

    \addplot[s-band,
    mark=square*,
    mark options={fill=white},
    only marks,
    each nth point=4,
    ] table[x=nspans,y=popt300theoryapprox] {data/HCF_Nspansweep.txt};
    
    \addplot[c-band,
    mark=square*,
    mark options={fill=white},
    only marks,
    each nth point=4,
    ] table[x=nspans,y=popt400theoryapprox] {data/HCF_Nspansweep.txt};

    \addplot[o-band,
    mark=*,
    mark options={fill=white},
    only marks,
    each nth point=4,
    ] table[x=nspans,y=popt100theory] {data/HCF_Nspansweep.txt};

    \addplot[e-band,
    mark=*,
    mark options={fill=white},
    only marks,
    each nth point=4,
    ] table[x=nspans,y=popt200theory] {data/HCF_Nspansweep.txt};

    \addplot[s-band,
    mark=*,
    mark options={fill=white},
    only marks,
    each nth point=4,
    ] table[x=nspans,y=popt300theory] {data/HCF_Nspansweep.txt};
    
    \addplot[c-band,
    mark=*,
    mark options={fill=white},
    only marks,
    each nth point=4,
    ] table[x=nspans,y=popt400theory] {data/HCF_Nspansweep.txt};
    
    \node at (rel axis cs: 0.25, 0.95) {\large (a)};

    \nextgroupplot[
    ymode = log,
    ymin=10,
    ymax=80,
    ytick={10,20,30,40,50,60,70,80},
    yticklabels={$10$,$20$,$30$,$40$,$50$,$60$,$70$,$80$},
    ylabel=Minimum $E_b$ (pJ/bit),
    ylabel shift = -5 pt,
    ]
    
    \addplot[o-band,
    no marks,
    line width = 1pt,
    ] table[x=nspans,y=eb100] {data/HCF_Nspansweep.txt};

    \addplot[e-band,
    no marks,
    line width = 1pt,
    ] table[x=nspans,y=eb200] {data/HCF_Nspansweep.txt};

    \addplot[s-band,
    no marks,
    line width = 1pt,
    ] table[x=nspans,y=eb300] {data/HCF_Nspansweep.txt};
    
    \addplot[c-band,
    no marks,
    line width = 1pt,
    ] table[x=nspans,y=eb400] {data/HCF_Nspansweep.txt};
    \legend{100 km, 200 km, 300 km, 400 km}

    \addplot[o-band,
    mark=square*,
    mark options={fill=white},
    only marks,
    each nth point=4,
    ] table[x=nspans,y=eb100theoryapprox] {data/HCF_Nspansweep.txt};

    \addplot[e-band,
    mark=square*,
    mark options={fill=white},
    only marks,
    each nth point=4,
    ] table[x=nspans,y=eb200theoryapprox] {data/HCF_Nspansweep.txt};

    \addplot[s-band,
    mark=square*,
    mark options={fill=white},
    only marks,
    each nth point=4,
    ] table[x=nspans,y=eb300theoryapprox] {data/HCF_Nspansweep.txt};
    
    \addplot[c-band,
    mark=square*,
    mark options={fill=white},
    only marks,
    each nth point=4,
    ] table[x=nspans,y=eb400theoryapprox] {data/HCF_Nspansweep.txt};

    \addplot[o-band,
    mark=*,
    mark options={fill=white},
    only marks,
    each nth point=4,
    ] table[x=nspans,y=eb100theory] {data/HCF_Nspansweep.txt};

    \addplot[e-band,
    mark=*,
    mark options={fill=white},
    only marks,
    each nth point=4,
    ] table[x=nspans,y=eb200theory] {data/HCF_Nspansweep.txt};

    \addplot[s-band,
    mark=*,
    mark options={fill=white},
    only marks,
    each nth point=4,
    ] table[x=nspans,y=eb300theory] {data/HCF_Nspansweep.txt};
    
    \addplot[c-band,
    mark=*,
    mark options={fill=white},
    only marks,
    each nth point=4,
    ] table[x=nspans,y=eb400theory] {data/HCF_Nspansweep.txt};

    \node at (rel axis cs: 0.1, 0.95) {\large (b)};
    
    \end{groupplot}

    \end{tikzpicture}

\caption{(a) Minimum-$E_b$ launch power and (b) minimum $E_b$ versus number of spans for different HCF span lengths. Solid lines indicate numerical results; circles are obtained from the closed-form solution in (\ref{eq:exactclosed}) and squares obtained from the approximation in (\ref{eq:approxclosed}).}
\label{fig:Nspan_hcf}
\end{figure}
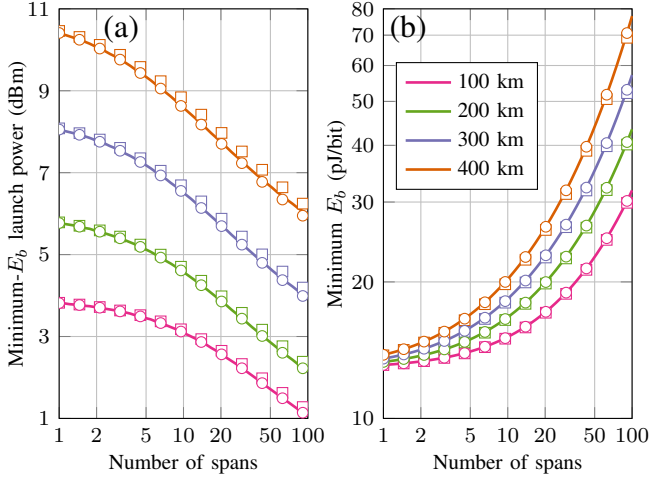

\pgfplotsset{compat=1.18}

\begin{figure*}[p]
\centering
\tikzsetnextfilename{contorSMF}
\begin{tikzpicture}[font=\footnotesize]
    \begin{groupplot}
    [
    width=0.5\linewidth,
    height=5.4cm,
    ylabel near ticks,
    xlabel near ticks,
    ylabel shift = -9 pt,
    xlabel shift = -2 pt,
    clip marker paths=true,
    xmode=log,
    ymode=log,
    xlabel=Attenuation (dB/km),
    ylabel=Per-span length (km),
    label style={font=\footnotesize},
    tick label style={font=\footnotesize},
    group style={group size=2 by 2,xlabels at=edge bottom,ylabels at=edge left,horizontal sep=0.7cm,vertical sep=1cm},
    title style={at={(0.5,0.95)},font=\bfseries\footnotesize},
    xmin=0.14,
    xmax=0.3,
    ymin=5,
    ymax=120,
    view={0}{90},
    colormap name=viridis,
    ytick = {5,10,20,40,70,120},
    yticklabels = {$5$,$10$,$20$,$40$,$70$,$120$},
    xtick = {0.14,0.16,0.2,0.25,0.3},
    xticklabels = {$0.14$,$0.16$,$0.2$,$0.25$,$0.3$},
    title style={yshift=-2pt},
    ]

    \def\dist{120pt}%

    \nextgroupplot[
    title = {\normalsize (a) Minimum $E_b$ (pJ/bit) at $P_{\text{opt}}^{E_b}$},
    ]
    
    \addplot3 [
        contour gnuplot={levels={16,17,18,20,25},
            contour label style={
            nodes={fill=white, inner sep=1pt},
            /pgf/number format/fixed,
            /pgf/number format/precision=1
            },
            label distance=\dist,
        },
        mesh/rows=25,
        mesh/cols=50,
    ] table {data/SMF_optEb_Eb.txt};

    \addplot[black,
    no marks,
    line width = 1pt,
    dashed,
    ] table{data/optline1_smf.txt};

    \node[red,fill=white,inner sep=1pt] at (axis cs:0.2,36) {(0.2, 45.6), 1.23~dBm/ch};
    \addplot[red,only marks,mark=*] coordinates {(0.2,45.616)};

    \node[red,fill=white,inner sep=1pt] at (axis cs:0.163,70) {(0.15, 58.9), 0.75~dBm/ch};
    \addplot[red,only marks,mark=triangle*,mark size=3pt] coordinates {(0.15,58.87)};

    \nextgroupplot[
    title = {\normalsize (b) Minimum $E_b$ (pJ/bit) at $P_{\text{opt}}^{T}$},
    ]
    
    \addplot3 [
        contour gnuplot={levels={16,17,18,20,25},
            contour label style={
            nodes={fill=white, inner sep=1pt},
            /pgf/number format/fixed,
            /pgf/number format/precision=1
            },
            label distance=\dist,
        },
        mesh/rows=25,
        mesh/cols=50,
    ] table {data/SMF_optcap_Eb.txt};

    \addplot[black,
    no marks,
    line width = 1pt,
    dashed,
    ] table{data/optline2_smf.txt};

    \node[red,red,fill=white,inner sep=1pt] at (axis cs:0.2,36) {(0.2, 45.6) 1.57~dBm/ch};
    \addplot[red,only marks,mark=*] coordinates {(0.2,45.616)};

    \node[red,red,fill=white,inner sep=1pt] at (axis cs:0.163,70) {(0.15, 58.9), 1~dBm/ch};
    \addplot[red,only marks,mark=triangle*,mark size=3pt] coordinates {(0.15,58.87)};

    \nextgroupplot[
    title = {\normalsize (c) Maximum $T$ (Tbps) at $P_{\text{opt}}^{E_b}$},
    ]
    
    \addplot3 [
        contour gnuplot={levels={25,35,42,44,45,46,47},
            contour label style={
            nodes={fill=white, inner sep=1pt},
            /pgf/number format/fixed,
            /pgf/number format/precision=1
            },
            label distance=\dist,
        },
        mesh/rows=25,
        mesh/cols=50,
    ] table {data/SMF_optEb_cap.txt};

    \addplot[black,
    no marks,
    line width = 1pt,
    dashed,
    ] table{data/optline3_smf.txt};

    \node[red,red,fill=white,inner sep=1pt] at (axis cs:0.2,23) {(0.2, 29.8) 0.09~dBm/ch};
    \addplot[red,only marks,mark=*] coordinates {(0.2,29.8176)};

    \node[red,red,fill=white,inner sep=1pt] at (axis cs:0.163,47) {(0.15, 40.2), -0.26~dBm/ch};
    \addplot[red,only marks,mark=triangle*,mark size=3pt] coordinates {(0.15,40.1535)};

    \nextgroupplot[
    title = {\normalsize (d) Maximum $T$ (Tbps) at $P_{\text{opt}}^{T}$},
    ]
    
    \addplot3 [
        contour gnuplot={levels={25,35,42,44,45,46,47},
            contour label style={
            nodes={fill=white, inner sep=1pt},
            /pgf/number format/fixed,
            /pgf/number format/precision=2
            },
            label distance=\dist,
        },
        mesh/rows=25,
        mesh/cols=50,
    ] table {data/SMF_optcap_cap.txt};

    \addplot[black,
    no marks,
    line width = 1pt,
    dashed,
    ] table{data/optline4_smf.txt};

    \node[red,red,fill=white,inner sep=1pt] at (axis cs:0.2,23) {(0.2, 30.1) 0.48~dBm/ch};
    \addplot[red,only marks,mark=*] coordinates {(0.2,30.13)};

    \node[red,red,fill=white,inner sep=1pt] at (axis cs:0.163,47) {(0.15, 40.2), 0.02~dBm/ch};
    \addplot[red,only marks,mark=triangle*,mark size=3pt] coordinates {(0.15,40.1535)};
    
    \end{groupplot}

    \end{tikzpicture}

\caption{Contour plot of SMF energy-per-bit $E_b$ (a, b) and throughput $T$ (c, d) for a 1000~km link as a function of per-span length and SMF attenuation at either minimum-$E_b$ (a, c) or maximum-$T$ (b, d) launch powers. Dashed line indicates the optimum span length achieving either minimum-$E_b$ or maximum-$T$ for a given fibre attenuation. Red markers indicate optimum span lengths and launch powers for different fibre attenuations.}
\label{fig:contour_smf}
\end{figure*}
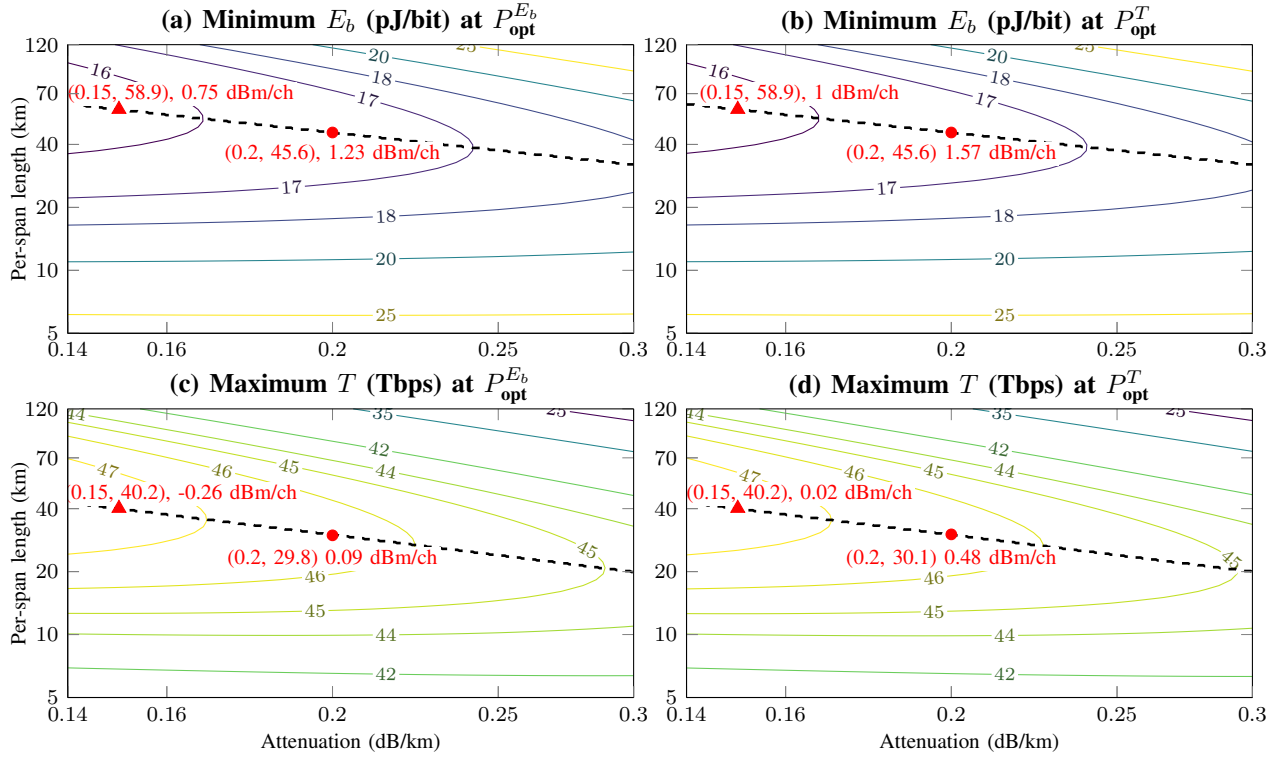
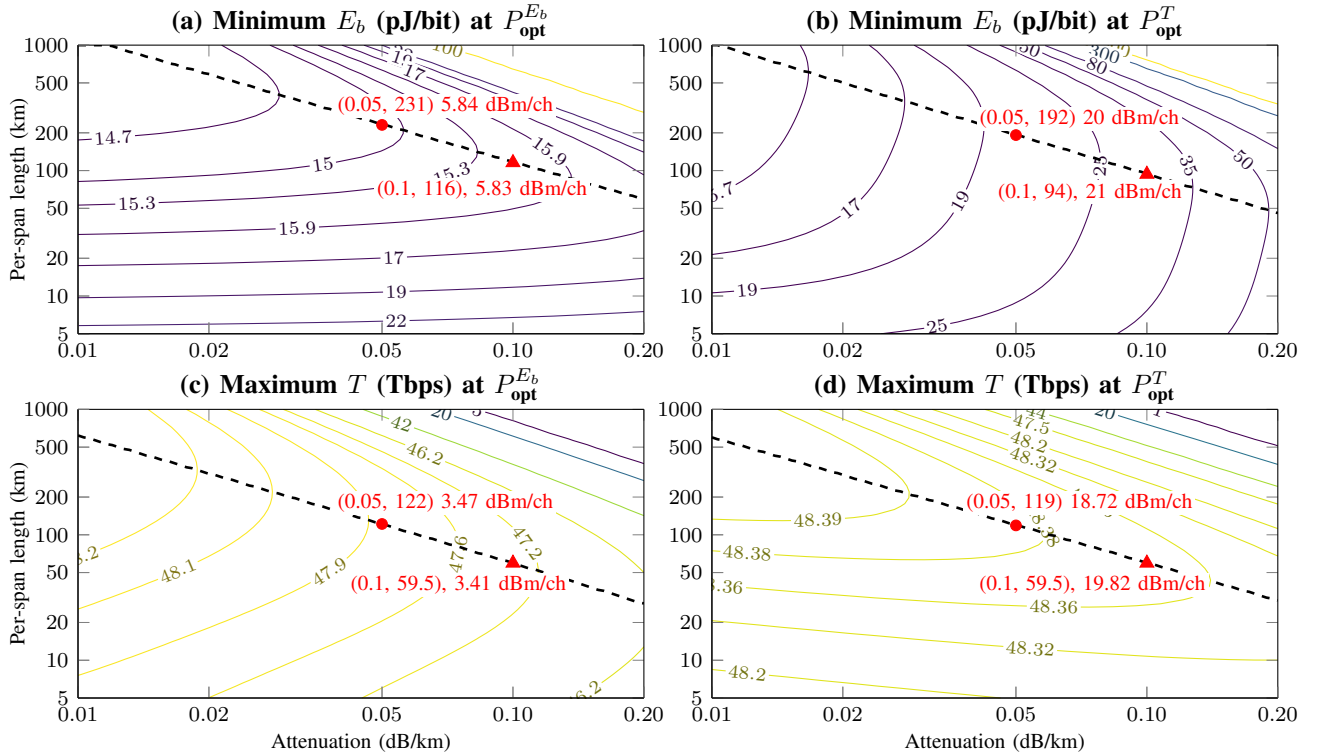
\begin{figure*}[p]
\centering
\tikzsetnextfilename{contorHCF}
\begin{tikzpicture}[font=\footnotesize]
    \begin{groupplot}
    [
    width=0.5\linewidth,
    height=5.4cm,
    ylabel near ticks,
    xlabel near ticks,
    ylabel shift = -9 pt,
    xlabel shift = -2 pt,
    clip marker paths=true,
    xmode=log,
    ymode=log,
    xlabel=Attenuation (dB/km),
    ylabel=Per-span length (km),
    clip mode=individual,
    label style={font=\footnotesize},
    tick label style={font=\footnotesize},
    group style={group size=2 by 2,xlabels at=edge bottom,ylabels at=edge left,horizontal sep=0.9cm,vertical sep=1cm},
    title style={at={(0.5,0.95)},font=\bfseries\footnotesize},
    xmin=0.01,
    xmax=0.2,
    ymin=5,
    ymax=1000,
    view={0}{90},
    colormap name=viridis,
    ytick = {5,10,20,50,100,200,500,1000},
    yticklabels = {$5$,$10$,$20$,$50$,$100$,$200$,$500$,$1000$},
    xtick = {0.01,0.02,0.05,0.10,0.20},
    xticklabels = {$0.01$,$0.02$,$0.05$,$0.10$,$0.20$},
    title style={yshift=-2pt},
    ]

    \def\dist{120pt}%

    \nextgroupplot[
    title = {\normalsize (a) Minimum $E_b$ (pJ/bit) at $P_{\text{opt}}^{E_b}$},
    ]
    
    \addplot3 [
        contour gnuplot={levels={14.7,15,15.3,15.9,17,19,22,100},
            contour label style={
            nodes={fill=white, inner sep=1pt},
            /pgf/number format/fixed,
            /pgf/number format/precision=1
            },
            label distance=\dist,
        },
        mesh/rows=25,
        mesh/cols=50,
    ] table {data/HCF_optEb_Eb.txt};

    \addplot[black,
    no marks,
    line width = 1pt,
    dashed,
    ] table{data/optline1_hcf.txt};

    \node[red,fill=white,inner sep=1pt] at (axis cs:0.07,330) {(0.05, 231) 5.84~dBm/ch};
    \addplot[red,only marks,mark=*] coordinates {(0.05,231.226)};
    
    \node[red,fill=white,inner sep=1pt] at (axis cs:0.085,70) {(0.1, 116), 5.83~dBm/ch};
    \addplot[red,only marks,mark=triangle*,mark size=3pt] coordinates {(0.1,115.718)};
    
    \nextgroupplot[
    title = {\normalsize (b) Minimum $E_b$ (pJ/bit) at $P_{\text{opt}}^{T}$},
    ]
    
    \addplot3 [
        contour gnuplot={levels={15.7,17,19,25,35,50,80,300,1000},
            contour label style={
            nodes={fill=white, inner sep=1pt},
            /pgf/number format/fixed,
            /pgf/number format/precision=1
            },
            label distance=90pt,
        },
        mesh/rows=25,
        mesh/cols=50,
    ] table {data/HCF_optcap_Eb.txt};

    \addplot[black,
    no marks,
    line width = 1pt,
    dashed,
    ] table{data/optline2_hcf.txt};

    \node[red,fill=white,inner sep=1pt] at (axis cs:0.07,260) {(0.05, 192) 20~dBm/ch};
    \addplot[red,only marks,mark=*] coordinates {(0.05,191.9)};
    
    \node[red,fill=white,inner sep=1pt] at (axis cs:0.074,67) {(0.1, 94), 21~dBm/ch};
    \addplot[red,only marks,mark=triangle*,mark size=3pt] coordinates {(0.1,93.518)};

    \nextgroupplot[
    title = {\normalsize (c) Maximum $T$ (Tbps) at $P_{\text{opt}}^{E_b}$},
    ]
    
    \addplot3 [
        contour gnuplot={levels={5,20,42,46.2,47.2,47.6,47.9,48.1,48.2},
            contour label style={
            nodes={fill=white, inner sep=1pt},
            /pgf/number format/fixed,
            /pgf/number format/precision=1
            },
            label distance=\dist,
        },
        mesh/rows=25,
        mesh/cols=50.5,
    ] table {data/HCF_optEb_cap.txt};

    \addplot[black,
    no marks,
    line width = 1pt,
    dashed,
    ] table{data/optline3_hcf.txt};

    \node[red,fill=white,inner sep=1pt] at (axis cs:0.07,180) {(0.05, 122) 3.47~dBm/ch};
    \addplot[red,only marks,mark=*] coordinates {(0.05,122)};
    
    \node[red,fill=white,inner sep=1pt] at (axis cs:0.075,40) {(0.1, 59.5), 3.41~dBm/ch};
    \addplot[red,only marks,mark=triangle*,mark size=3pt] coordinates {(0.1,59.5)};

    \nextgroupplot[
    title = {\normalsize (d) Maximum $T$ (Tbps) at $P_{\text{opt}}^{T}$},
    ]
    
    \addplot3 [
        contour gnuplot={levels={1,20,44,47.5,48.2,48.32,48.36,48.38,48.39},
            contour label style={
            nodes={fill=white, inner sep=1pt},
            /pgf/number format/fixed,
            /pgf/number format/precision=2
            },
            label distance=\dist,
        },
        mesh/rows=25,
        mesh/cols=50,
    point meta max=50.5,
    ] table {data/HCF_optcap_cap.txt};

    \addplot[black,
    no marks,
    line width = 1pt,
    dashed,
    ] table{data/optline4_hcf.txt};

    \node[red,fill=white,inner sep=1pt] at (axis cs:0.07,180) {(0.05, 119) 18.72~dBm/ch};
    \addplot[red,only marks,mark=*] coordinates {(0.05,118.84)};
    
    \node[red,fill=white,inner sep=1pt] at (axis cs:0.075,40) {(0.1, 59.5), 19.82~dBm/ch};
    \addplot[red,only marks,mark=triangle*,mark size=3pt] coordinates {(0.1,59.5)};
    
    \end{groupplot}

    \end{tikzpicture}

\caption{Contour plot of HCF energy-per-bit $E_b$ (a, b) and throughput $T$ (c, d) for a 1000~km link as a function of per-span length and HCF attenuation at either minimum-$E_b$ (a, c) or maximum-$T$ (b, d) launch powers. Dashed line indicates the optimum span length achieving either minimum-$E_b$ or maximum-$T$ for a given fibre attenuation. Red markers indicate optimum span lengths and launch powers for different fibre attenuations.}
\label{fig:contour_hcf}
\end{figure*}

By making further approximations in (\ref{eq:exactclosed}) to eliminate small terms, the approximated closed-form solution is

\begin{equation} \label{eq:approxclosed}
    P_{\text{opt}}^{E_b} \approx \sqrt{ \frac{\eta_{\text{PCE}}P_{\text{ASE}} (N_{\text{ch}}P_{\text{TRX}} + N_{\text{span}}P_{\text{mm}})  }{N_{\text{ch}}N_{\text{span}}(1-1/G)\kappa (1+\kappa) \ln(1+1/\kappa)} }
\end{equation}

While (\ref{eq:approxclosed}) is a more approximate solution than (\ref{eq:exactclosed}), it provides some general insight. Firstly, $ \lim_{\kappa \to 0} P_{\text{opt}}^{E_b} \to +\infty$, thus an $E_b$ minimum only exists when transceiver noise or IMI are present. Note that nonlinearity is neglected in the closed-form here, so $P_{\text{opt}}^{E_b}$ is determined by the smaller of (\ref{eq:maxthroughput}) and (\ref{eq:exactclosed})/(\ref{eq:approxclosed}), but for HCF the latter terms are typically more than an order of magnitude smaller. The minimum-$E_b$ launch power is directly proportional to the PCE, ASE noise power and the combined launch-power-independent overheads ($N_{\text{ch}}P_{\text{TRX}} + N_{\text{span}}P_{\text{mm}}$). This can then be plugged into (\ref{eq:generalform}) to determine the minimum $E_b$.

Fig.~\ref{fig:Nspan_hcf} shows the verification of the closed-form solutions given in (\ref{eq:exactclosed}) and (\ref{eq:approxclosed}), similar to Fig.~\ref{fig:Nspan_smf}. 
The closed-form shows good agreement with the numerical simulation (solid line), with (\ref{eq:exactclosed}) showing better agreement than (\ref{eq:approxclosed}) as the number of spans is increased. Since $E_b$ for HCF has low curvature near $P_{\text{opt}}^{E_b}$, a small error in launch power introduces only a nominal relative error on the minimum $E_b$.
For 10-spans the $P_{\text{opt}}^{E_b}$ error is 0.2\% (Eq.~\ref{eq:exactclosed}) and 4.75\% (Eq.~\ref{eq:approxclosed}) for the worst-case 400~km span length, corresponding to $E_b$ error of less than $10^{-4}$\% (Eq.~\ref{eq:exactclosed}) and 1.2\% (Eq.~\ref{eq:approxclosed}).

\subsection{Optimum span lengths in $P_{\text{opt}}^{E_b}$ and $P_{\text{opt}}^{T}$ regimes}

Using the numerical simulation to solve (\ref{eq:generalform}), the optimum span length can be investigated for both the minimum-$E_b$ and maximum-$T$ launch power regimes for both SMF and HCF systems. The parameters used in this simulation are the same as those described in Section~\ref{sec:methodology}.

Fig.~\ref{fig:contour_smf}(a-d) shows contour plots of the energy per bit and throughput as a function of span length and SMF attenuation for a 1000~km link, calculated at either $P_{\text{opt}}^{E_b}$ or at $P_{\text{opt}}^{T}$. It is immediately clear that Fig.~\ref{fig:contour_smf}(a) and Fig.~\ref{fig:contour_smf}(b) $E_b$ surfaces are identical, Fig.~\ref{fig:contour_smf}(c) and Fig.~\ref{fig:contour_smf}(d) also only differ marginally. The optimum launch powers in all four cases are also very similar within 2~dB of each other. This is in agreement with our previous observation for SMF, $P_{\text{opt}}^{E_b} \approx P_{\text{opt}}^{T}$, thus maximising throughput will also minimise energy per bit in SMF. Comparing the red markers between Fig.~\ref{fig:contour_smf}(a-b) with Fig.~\ref{fig:contour_smf}(c-d), it can also be noted that the optimum span length that minimises $E_b$ is not the same as that which maximises $T$. For 0.15~dB/km and 0.2~dB/km attenuation, the minimum-$E_b$ span lengths are 58.9~km and 45.6~km, and the maximum-$T$ span lengths are 40.2~km and 29.8~km respectively.

Figs.~\ref{fig:contour_hcf}(a-d) shows the same plot as Fig.~\ref{fig:contour_smf} for a 1000~km HCF system. Numerical simulations were used here to ensure accuracy of the plot compared to the closed form in edge cases, e.g. when the span length >500~km. The difference between Fig.~\ref{fig:contour_smf} and Fig.~\ref{fig:contour_hcf} is apparent; the $P_{\text{opt}}^{E_b}$ and $P_{\text{opt}}^{T}$ contour surfaces are no longer identical. This highlights the observation from the previous section, that in HCF $P_{\text{opt}}^{E_b} \neq P_{\text{opt}}^{T}$. The optimal launch powers deviate significantly between scenarios, with the $P_{\text{opt}}^{T}$ case (Figs.~\ref{fig:contour_hcf}(b,d)) in the range 18 to 20~dBm, and the $P_{\text{opt}}^{E_b}$ case (Figs.~\ref{fig:contour_hcf}(a,c)) below 6~dBm. Operating at $P_{\text{opt}}^{E_b}$leads to significant reduction in output power, with only marginal reduction in overall throughput.

For 0.05~dB/km and 0.1~dB/km attenuation, the minimum-$E_b$ span lengths for $P_{\text{opt}}^{E_b}$ are 231~km and 116~km, and the maximum-$T$ span lengths for $P_{\text{opt}}^{T}$ are 119~km and 59.5~km respectively. While the optimum span length depends significantly on the parameters used (e.g. $\eta_{\text{PCE}},P_{\text{TRX}}$ etc), Fig.~\ref{fig:contour_hcf} shows that energy efficiency considerations can significantly alter the design of HCF transmission systems.

\section{HCF vs SMF comparison for a 1000~km link}

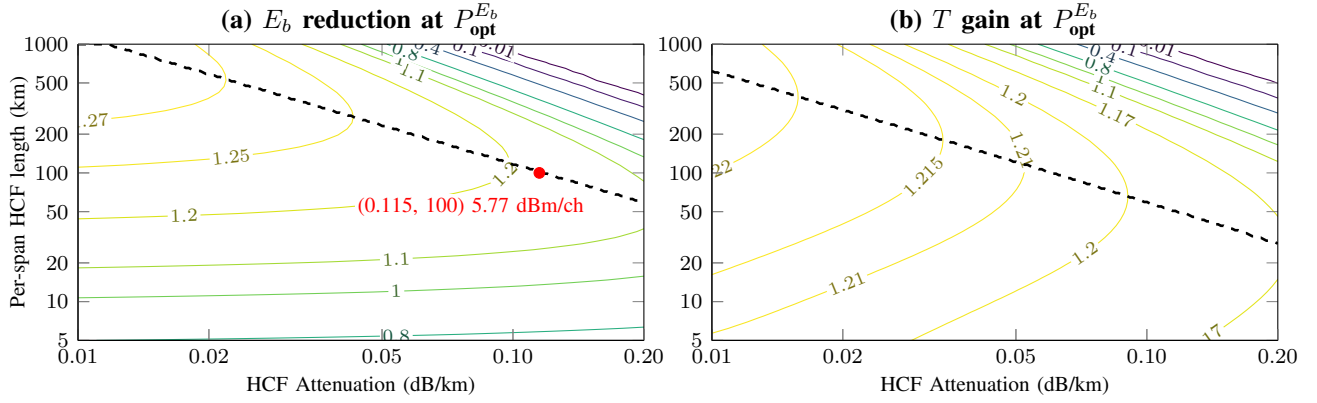
\begin{figure*}[t]
\centering
\tikzsetnextfilename{contorHCFvsSMF}
\begin{tikzpicture}[font=\footnotesize]
    \begin{groupplot}
    [
    width=0.5\linewidth,
    height=5.5cm,
    ylabel near ticks,
    xlabel near ticks,
    ylabel shift = -9 pt,
    xlabel shift = -2 pt,
    clip marker paths=true,
    xmode=log,
    ymode=log,
    xlabel=HCF Attenuation (dB/km),
    ylabel=Per-span HCF length (km),
    clip mode=individual,
    label style={font=\footnotesize},
    tick label style={font=\footnotesize},
    group style={group size=2 by 1,xlabels at=edge bottom,ylabels at=edge left,horizontal sep=0.9cm},
    title style={at={(0.5,0.95)},font=\bfseries\footnotesize},
    xmin=0.01,
    xmax=0.2,
    ymin=5,
    ymax=1000,
    view={0}{90},
    colormap name=viridis,
    ytick = {5,10,20,50,100,200,500,1000},
    yticklabels = {$5$,$10$,$20$,$50$,$100$,$200$,$500$,$1000$},
    xtick = {0.01,0.02,0.05,0.10,0.20},
    xticklabels = {$0.01$,$0.02$,$0.05$,$0.10$,$0.20$},
    title style={yshift=-2pt},
    ]

    \def\dist{120pt}%

    \nextgroupplot[
    title = {\normalsize (a) $E_b$ reduction at $P_{\text{opt}}^{E_b}$},
    ]
    
    \addplot3 [
        contour gnuplot={levels={0.01,0.1,0.4,0.8,1,1.1,1.2,1.25,1.27},
            contour label style={
            nodes={fill=white, inner sep=1pt},
            /pgf/number format/fixed,
            /pgf/number format/precision=3
            },
            label distance=\dist,
        },
        mesh/rows=25,
        mesh/cols=50,
    ] table {data/SMFvsHCF_optEb.txt};

    \addplot[black,
    no marks,
    line width = 1pt,
    dashed,
    ] table{data/optline1_hcfvssmf.txt};

    \node[red,fill=white,inner sep=1pt] at (axis cs:0.08,55) {(0.115, 100) 5.77~dBm/ch};
    \addplot[red,only marks,mark=*] coordinates {(0.115,100)};

    \nextgroupplot[
    title = {\normalsize (b) $T$ gain at $P_{\text{opt}}^{E_b}$},
    ]
    
    \addplot3 [
        contour gnuplot={levels={0.01,0.1,0.4,0.8,1,1.1,1.17,1.2,1.21,1.215,1.22},
            contour label style={
            nodes={fill=white, inner sep=1pt},
            /pgf/number format/fixed,
            /pgf/number format/precision=3
            },
            label distance=90pt,
        },
        mesh/rows=25,
        mesh/cols=50,
    ] table {data/SMFvsHCF_optEb_cap.txt};

    \addplot[black,
    no marks,
    line width = 1pt,
    dashed,
    ] table{data/optline2_hcfvssmf.txt};

    \end{groupplot}

    \end{tikzpicture}

\caption{Contour plot of HCF gain over a 10x100~km SMF link in terms of energy-per-bit $E_b$ (a) and throughput $T$ (b) at minimum-$E_b$ launch power for a fixed 1000~km HCF length. Dashed lines indicates the optimum span length achieving maximum HCF gains. Red marker indicates optimum attenuation and launch power for 100~km HCF span length.}
\label{fig:contour_hcfvssmf}
\end{figure*}

Fig.~\ref{fig:contour_hcfvssmf} shows the HCF gain in terms of energy per bit and throughput over a 10x100km (1000~km total) SMF link with the same fibre parameters as Section~\ref{sec:methodology}. The total HCF link length was 1000~km, the same as the SMF reference link, thus a per-span length of 100~km means both the SMF and HCF links have the same number of repeaters. 
Operating at $P_{\text{opt}}^{E_b}$ in both fibres (for SMF this is the same as $P_{\text{opt}}^{T}$) the HCF reduction in $E_b$ is more than 1.2$\times$ using HCF with attenuation less than 0.1~dB/km and link lengths above 50~km. This is also true for the throughput, achieving above 1.2$\times$ throughput improvement for almost all attenuation and per-span lengths (except where the HCF is extremely long and lossy). Assuming the 100~km SMF is directly replaced with 100~km HCF, the maximum $E_b$ reduction occurs at 0.115~dB/km attenuation, achieving 1.18$\times$ improvement in both $E_b$ and $T$. It is therefore possible to maintain the HCF's throughput advantages over SMF whilst operating at $P_{\text{opt}}^{E_b}$ to maximise energy efficiency.

\section{Energy efficiency gains in HCF links}

Table~\ref{tab:comparison} shows the change in throughput and energy per bit for different number of spans and span lengths when moving from 
$P_{\text{opt}}^{T}$ to $P_{\text{opt}}^{E_b}$ within a HCF system. For very short links (i.e. consisting of few spans) the gains are relatively small, only a few percent reduction in $E_b$ with nominal throughput penalty. However, as the number of spans increases, particularly when the total link length is large, the gains become more significant. For 10x300km transmission, $E_b$ is reduced by more than 50\% for only a 3.3\% throughput penalty. Also note that the total launch power when operating at $P_{\text{opt}}^{E_b}$ per channel is approximately 20~dBm, whereas in $P_{\text{opt}}^{T}$ operation it is more than 33~dBm. This leads to an interesting observation regarding amplifier design - energy-efficient HCF transmission systems do not require novel high-power amplifiers, since the total $P_{\text{opt}}^{E_b}$ values are well within the achievable outputs of modern EDFAs. This eases the complexity of HCF deployment, since existing EDFAs can be re-used for these systems, and also reduces the risk of component failure as well as link optical safety requirements associated with high-power amplification in field deployment \cite{Yang2026field}.

\begin{table}
\centering
\begin{tabulary}{\columnwidth}{@{}CCCCCC@{}}
\toprule
Span length (km) & $N_{\text{span}}$ & Total $P_{\text{opt}}^{E_b}$ (dBm) & Total $P_{\text{opt}}^{T}$ (dBm) &  $\Delta T$ (\%) & $\Delta E_b$ (\%) \\ \midrule
    & 1  & 18.4 & 33   & -0.13 & -3.5  \\
100 & 5  & 18.0 & 33   & -0.59 & -15.2 \\
    & 10 & 17.7 & 33   & -1.07 & -26.0 \\ \midrule
    & 1  & 20.4 & 34.7 & -0.27 & -6.60 \\
200 & 5  & 19.7 & 34.7 & -1.15 & -25.7 \\
    & 10 & 19.2 & 34.7 & -1.98 & -40.3 \\ \midrule
    & 1  & 22.7 & 36.4 & -0.48 & -10.0 \\
300 & 5  & 21.8 & 36.4 & -1.96 & -35   \\
    & 10 & 21.1 & 36.4 & -3.3  & -51   \\ \bottomrule
\end{tabulary}%
\caption{Throughput and energy-per-bit change when moving from $P_{\text{opt}}^{T}$ to $P_{\text{opt}}^{E_b}$ operating points for different HCF transmission scenarios for 0.05~dB/km attenuation.}
\label{tab:comparison}
\end{table}
\section{Conclusion} \label{sec:conclusion}

In this work, we investigated the energy efficiency of coherent HCF transmission systems. We showed that, unlike SMF, HCF links have separate optimum launch powers for either minimum-energy-per-bit or maximum-throughput. We derived a closed form equation for the minimum-energy-per-bit launch power, showing excellent agreement with the numerical simulation. 
We showed that energy-efficient link optimisation is considerably different in HCF systems versus SMF, with energy per bit maximisation and throughput maximisation each leading to different optimum span lengths and launch powers in HCF. Our results showed that it is not sufficient to consider only throughput, since it obscures significant trade-offs in terms of energy per bit. The low nonlinearity of HCF leads to extremely high launch powers (\textgreater 33~dBm) to achieve throughput maximisation, but the benefits are marginal due to the transceiver- and IMI-noise limited SNR. This limitation only becomes exacerbated as HCF loss decreases and the span lengths increase. Operating at the most energy-efficient point not only minimises $E_b$, but reduces the requirements on the amplifier hardware, enabling existing SMF-class EDFAs to be used in hollow-core fibre systems.

\appendix
To find the minimum of (\ref{eq:rearrange}), the substitution
\begin{equation} \label{eq:sub}
    \delta = \frac{P_{\text{ASE}}}{P_{\text{ASE}} + \kappa P}
\end{equation}

was used, where $\delta$ is the ratio of ASE power to the total noise power. Taking the derivative of (\ref{eq:rearrange}) with respect to $P$, this leads to a transcendental equation in $\delta$
\begin{equation} \label{eq:appendixquadratic}
    (1 - r)\delta^2 - \delta + (1+\kappa-\delta)\ln\Big(1+\frac{1-\delta}{\kappa}\Big) = 0
\end{equation}

where $r$ is given in (\ref{eq:r}). The log term can be rewritten as
\begin{equation} \label{eq:appendixapprox}
    \ln(1+1/\kappa) + \ln\Big(1-\frac{\delta}{\kappa+1}\Big) \approx \ln(1+1/\kappa) -\frac{\delta}{\kappa+1}
\end{equation}

where the approximation $\ln(1-x) \approx -x$ for $x \ll 1$, therefore $\delta \ll \kappa+1$. This implies that the ASE noise power must be small compared to the transceiver noise power at $P_{\text{opt}}^{E_b}$, which becomes less accurate for very long, lossy fibre spans (since $P_{\text{ASE}}$ scales with the amplifier gain), or when $P_{\text{opt}}^{E_b}$ is at a very small launch power.

\ifCLASSOPTIONcaptionsoff
  \newpage
\fi
\section*{Acknowledgement}
For the purpose of open access, the author has applied a Creative Commons Attribution (CC BY) licence to any Author Accepted Manuscript version arising.

\bibliographystyle{IEEEtran}
\bibliography{IEEEabrv,mybib}

\end{document}